\documentclass[12pt,times]{article}

 \usepackage[pdftex]{color,graphicx}
\usepackage{enumerate}
\usepackage{amsmath}
\usepackage{amssymb}
\usepackage{alltt}
\usepackage{setspace}
\usepackage{sectsty}
\partfont{\small}
\sectionfont{\small}
\subsectionfont{\small}
\usepackage[left=2cm,top=2cm,right=3cm,head=1cm,foot=1cm]{geometry}
\numberwithin{equation}{section}
\let\oldsqrt\sqrt
\def\sqrt{\mathpalette\DHLhksqrt}
\def\DHLhksqrt#1#2{%
\setbox0=\hbox{$#1\oldsqrt{#2\,}$}\dimen0=\ht0
\advance\dimen0-0.2\ht0
\setbox2=\hbox{\vrule height\ht0 depth -\dimen0}%
{\box0\lower0.4pt\box2}}
\newcommand{\al}{\alpha}
\newcommand{\g}{\gamma}
\newcommand{\e}{\varepsilon}
\newcommand{\ta}{\theta}
\newcommand{\z}{\zeta}
\newcommand{\G}{\Gamma}

\newcommand{\ph}{\varphi}
\newcommand{\da}{\dagger}

\newcommand{\la}{\mathcal L}

\newcommand{\Ld}{\Lambda}

\newcommand{\M}{\mathcal M}

\newcommand{\+}{\oplus}
\newcommand{\x}{\otimes}
\newcommand{\ml}{\left(\begin{matrix}}
\newcommand{\mr}{\end{matrix}\right)}

\newcommand{\w}{\omega}

\newcommand{\tr}{\text{tr}}
\newcommand{\op}{\mathcal O}
\newcommand{\del}{\delta}

\newcommand{\lrarrow}{\leftrightarrow}

\newcommand{\ka}{\kappa}

\begin{document}
\vspace*{-2cm}
\begin{center}\textbf{Neutrino Mixing and the Double Tetrahedral Group}\end{center}
\begin{center} Yoni BenTov$^1$ and A. Zee$^{1,2}$\end{center}
$^1\,$\textit{Department of Physics, University of California, Santa Barbara CA 93106}\\
$^2\,$\textit{Kavli Institute for Theoretical Physics, University of California, Santa Barbara CA 93106}\\
\begin{center}\textbf{Abstract}\end{center}
In the spirit of a previous study of the tetrahedral group $T \simeq A_4$, we discuss a minimalist scheme to derive the neutrino mixing matrix using the double tetrahedral group $T'$, the double cover of $T$. The new features are three distinct 2-dimensional representations and complex Clebsch-Gordan coefficients, which can result in a geometric source of CP violation in the neutrino mass matrix. In an appendix, we derive explicitly the relevant group theory for the tetrahedral group $T$ and its double cover $T'$.
%\\\\
\section{Neutrino Mixing Matrix}\label{sec:intro}
%\textbf{I. Neutrino Mixing Matrix.}
%\\\\
The neutrino mixing matrix $V$ is defined as the transformation matrix that takes the neutrino mass eigenstates to the charged lepton mass eigenstates:
\[\text{charged lepton mass basis}\rightarrow\ml \nu_e\\ \nu_\mu \\ \nu_\tau \mr = V\,\ml \nu_1\\ \nu_2 \\ \nu_3 \mr\leftarrow\text{neutrino mass basis}\;.\]
We assume that the neutrino sector consists of 3 light Majorana neutrinos, so that $V$ is a 3-by-3 unitary matrix. Under this assumption, the present bounds on $V$ are \cite{review}\\
 \begin{equation}\label{eq:Vexp}
 |V_{\text{exp}}| \approx \ml 0.77\!\!-\!0.86&0.50\!\!-\!0.63&0.00\!\!-\!0.22\\0.22\!\!-\!0.56&0.44\!\!-\!0.73&0.57\!\!-\!0.80\\0.21\!\!-\!0.55&0.40\!\!-\!0.71&0.59\!\!-\!0.82 \mr\;.
 \end{equation}
The notation $|V|$ is such that the element in the $\al^{\text{th}}$ row and $i^{\text{th}}$ column of the matrix $|V|$ is the absolute value of $V_{\al i}$ (with $\al = e,\mu,\tau$ and $i = 1,2,3$). An arbitrary 3-by-3 unitary matrix $V$ can be parameterized as \cite{KM, jarlskog}
\begin{equation}\label{eq:KPM}
V = \ml e^{\,i\kappa_1}&0&0\\0&e^{\,i\kappa_2}&0\\0&0&e^{\,i\kappa_3} \mr\ml -c_2c_3&c_2s_3&\hat s_2^*\\ c_1s_3+s_1\hat s_2c_3&c_1c_3-s_1\hat s_2s_3&s_1c_2\\ s_1s_3-c_1\hat s_2c_3 & s_1c_3+c_1\hat s_2s_3 & -c_1c_2 \mr \ml e^{\,i\rho}&0&0\\0&e^{\,i\sigma}&0\\0&0&1 \mr
\end{equation}
where $c_I \equiv \cos\ta_I$, $s_I \equiv \sin\ta_I$ and $\hat s_2 \equiv s_2\,e^{\,i\delta_{\text{CP}}}$. The phase angles $\kappa_i$ can be absorbed into overall rephasing of the charged lepton fields, and so from now on we set them equal to zero. The phase angles $\rho$ and $\sigma$ are physically meaningful parameters that violate CP, but they are not observable in neutrino oscillations. The phase angle $\delta_{\text{CP}}$ is also a physically meaningful parameter that violates CP, but unlike $\rho$ and $\sigma$ it is in principle observable in oscillations. 
\\
\pagebreak
\\
Neutrino oscillations also give us the mass-squared differences $m_{21}^2 \equiv m_2^2-m_1^2$ and $m_{31}^2 \equiv m_3^2-m_1^2$, where $m_i$ is the physical mass of the $i^{\text{th}}$ neutrino. The present data \cite{review} give $m_{21}^2 = 7.67_{-0.61}^{+0.67}\times10^{-5}$ eV$^2$ and  
\[m_{31}^2 = \left\{ \begin{matrix} -2.37_{-0.46}^{+0.43}\times10^{-3}\text{eV}^2\;\;\text{(inverted hierarchy)}\\\\ +2.46_{-0.42}^{+0.47}\times10^{-3}\text{eV}^2\;\;\text{(normal hierarchy)}  \end{matrix} \right.\]
\\
which imply that the ratio $R \equiv m_{31}^2/m_{21}^2$ is roughly in the range $-34 < R < -27$ for the inverted case and $+24 < R < +42$ for the normal case. When we attempt to construct a theory of neutrino oscillations, the data points we seek to explain are $|V_{\al i}|$ and $R$.
\\\\
Suppose that the entries in $V$ arise from a discrete flavor symmetry that predicts a Lagrangian of the form\footnote{We use two-component spinor notation, for which the Lorentz-spinor indices $\al = 1,2$ contract as $\nu\nu \equiv \nu^\al\nu_\al \equiv \e_{\al\beta}\nu^\al\nu^\beta$, so that $\nu\nu' = \nu'\nu$ for two Grassmann-valued fields $\nu_\al$ and $\nu'_\al$. For a thorough review of this notation, see \cite{2spinor}.}
\[\la = - \vec e^{\,T}\,M_\ell\,\vec e^{\,c} - \tfrac{1}{2}\vec\nu^{\,T}\,M_\nu\,\vec\nu + h.c.\]
with $M_\ell$ and $M_\nu$ being generic 3-by-3 complex matrices\footnote{Since we assume the neutrinos to be Majorana, the matrix $M_\nu$ will always be symmetric.}. Then define the charged lepton mass basis by $\vec e \equiv U_L \vec e_{\ell\text{-mass}}$ and $\vec e^{\,c} \equiv U_R \vec e^{\,c}_{\ell\text{-mass}}$ such that $U_L^T M_\ell U_R \equiv D_\ell$, where $D_\ell \equiv \text{diag}(m_e,m_\mu,m_\tau)$ with $m_e,m_\mu,m_\tau$ real and positive. Similarly, define the neutrino mass basis by $\vec\nu \equiv U_\nu \vec\nu_{\nu\text{-mass}}$ such that $U_\nu^TM_\nu U_\nu \equiv D_\nu$, where $D_\nu \equiv \text{diag}(m_1,m_2,m_3)$ with $m_1,m_2,m_3$ real and positive \cite{hahn}. To summarize, we have
\[\ml \vec \nu\\ \vec e \mr = \ml U_\nu \vec \nu_{\nu\text{-mass}}\\ U_L \vec e_{\ell\text{-mass}} \mr = U_L\ml U_L^{-1}U_\nu \vec \nu_{\nu \text{-mass}}\\ \vec e_{\ell\text{-mass}} \mr\;.\]
The unitary matrix $V \equiv U_L^{-1}U_\nu$ is the neutrino mixing matrix.
%\\\\
%\textbf{II. A Minimalist Framework.}
%\\\\
\section{A Minimalist Framework}\label{sec:minimal}
In the interest of adding as little theoretical structure as possible to explain neutrino mixing, we stick to the following rules. First, we assume only the minimal fermion content of the Standard Model. Second, we extend the Higgs sector using only scalars that transform as doublets under the electroweak $SU(2)_\text{W}\x U(1)_\text{Y}$ gauge group (rather than singlets, triplets or higher representations). Third, we assume as in the Standard Model that the charged lepton masses come from dimension-4 operators of the form $\la \sim \e^{IJ}\ph_I\ell_J e^c+h.c.$, where $\ell \equiv \ml \nu\\ e \mr$ denotes the usual lepton doublet, $I = 1,2$ and $J = 1,2$ are $SU(2)_{\text{W}}$ indices (which we suppress in all subsequent sections), and $\ph$ denotes generically any $SU(2)_{\text{W}}\x U(1)_{\text{Y}}$ Higgs doublet. Fourth, we assume that lepton number is broken so that the neutrinos gain Majorana masses through dimension-5 operators of the form $\la \sim \tilde\ph^{\da I}\tilde\ph'^{\da J}\ell_I\ell_J+h.c.$, where $\tilde\ph$ and $\tilde\ph'$ denote generically any Higgs doublets that may or may not be the same as $\ph$. 
\\\\
In the spirit of treating our $T'$-based theories as modules to be embedded in a larger structure, we will implicitly assume that extra restrictions exist which forbid the Higgs fields in the charged lepton sector from coupling to the Higgs fields in the neutrino sector, and vice versa. For notational convenience, we will denote the vacuum expectation value of the neutral component of a Higgs field simply by the name of the field.
%\\\\
%\textbf{III. Double Tetrahedral Group.}
%\\\\
\section{Double Tetrahedral Group}\label{sec:doubleT}
For years, Ma \cite{ma} and many others have advocated the use of the tetrahedral group $T \simeq A_4$ to explain the large neutrino mixing angles. Other authors \cite{Tprime} have further turned to the double tetrahedral group $T'$ in order to separate one family of fermions from the other two. Of particular interest is an $SU(5)$ grand unified model proposed by Chen and Mahanthappa in which CP violation arises from the complex Clebsch-Gordan coefficients of $T'$ \cite{chen}. 
\\\\
We will now explain these remarks and propose an effective field theory that does not rely on a particular high-energy completion. In a previous paper \cite{zeeA4}, one of us has given a pedagogical introduction to the tetrahedral group, which we now summarize briefly\footnote{For the convenience of the reader, in the body of the paper we quote various results that are derived in the appendix.} as a preface to the double tetrahedral group. The tetrahedral group $T$ is the group whose 3-dimensional representation is given by the collection of rotations that leave a regular tetrahedron invariant. The group also has three distinct singlet representations, denoted by $1,1'$ and $1''$. To build models we need representation multiplication rules and the associated Clebsch-Gordan coefficients. Given two triplets $v = (v_1,v_2,v_3) \sim 3$ and $w = (w_1,w_2,w_3) \sim 3$ of the group $T$, we have the multiplication rule $3\x3 = 1\+1'\+1''\+3_1\+3_2$, where\footnote{The subscript $n$ in $(vw)_n$ denotes the particular irreducible representation $n$ on the right-hand side of the equation $``3\x3 = n\+..."$, whereas the subscripts on terms such as $``v_1w_1+\w\,v_2w_3+\w^*v_3w_3"$ denote components $v_i$ and $w_i$ of the triplets $\vec v$ and $\vec w$.}:
\begin{align*}
&(vw)_1 = v_1w_1+v_2w_2+v_3w_3\\
&(vw)_{1'} = v_1w_1+\w\,v_2w_2+\w^*v_3w_3\\
&(vw)_{1''} = v_1w_1+\w^*v_2w_2+\w\,v_3w_3\\
&(vw)_{3_1} = (v_2w_3,v_3w_1,v_1w_2)\\
&(vw)_{3_2} = (v_3w_2,v_1w_3,v_2w_1)\;.
\end{align*}
The complex phase $\w \equiv e^{\,i2\pi/3}$ satisfies $1+\w+\w^* = 0$. Although the $1'$ and $1''$ are invariant under the rotations through $\pi$, they transform as $(vw)_{1'} \to \w\,(vw)_{1'}$ and $(vw)_{1''} \to \w^*(vw)_{1''}$ under the cyclic permutation $123\to312$. This also implies the multiplication rule $1^a\x 1^b = 1^{a+b}$ for the singlets, where $a$ and $b$ can be 0, 1, or 2 to denote the number of primes mod 3. This is all we need to construct theories of neutrino mixing based on tetrahedral symmetry. 
\pagebreak\\
The tetrahedral group is a subgroup of $SO(3)$, and $SO(3)$ is locally isomorphic to $SU(2)$. We may thereby define a subgroup of $SU(2)$ whose image in $SO(3)$ leaves invariant a regular tetrahedron. Since $SU(2)$ double covers $SO(3)$, the result is the double cover of the tetrahedral group $T$, which is called the double tetrahedral group $T'$ \cite{subgroups}. In practical terms, the new feature of this construction is that we gain three distinct 2-dimensional representations, denoted by $2,2'$ and $2''$. 
\\\\
Given two $T'$-doublets $\chi = (\chi_1,\chi_2) \sim 2$ and $\xi = (\xi_1,\xi_2) \sim 2$, we have the multiplication rule $2\x2 = 1\+3$ familiar from $SU(2)$, where the invariant singlet is $ (\chi\xi)_1 = \e^{ij}\chi_i\xi_j = \chi_1\xi_2-\chi_2\xi_1 \sim 1$, and the triplet is\footnote{For a derivation, see the discussion leading to Eq. (\ref{eq:2times2}).}
\[(\chi\xi)_3 = \ml -i(\chi_1\xi_1+\chi_2\xi_2)\\ -(\chi_1\xi_1-\chi_2\xi_2)\\ \chi_1\xi_2+\chi_2\xi_1 \mr\;.\]
Just as we had the rule $1^a\x 1^b = 1^{a+b}$, we now also have the rule $2^a\x2^b = 1^{a+b}\+3$. To understand what that means, consider the doublet $\chi = (\chi_1,\chi_2) \sim 2$ again, but this time consider a doublet $\xi' = (\chi_1',\chi_2') \sim 2'$. The product $(\chi\xi') \sim 2\x2' = 1'\+3$ contains the singlet $(\chi\xi')_{1'} = \chi_1\xi'_2-\chi_2\xi'_1 \sim 1'$, which as explained previously is not invariant under $T'$, and the triplet\footnote{For a derivation, see the discussion leading to Eqs. (\ref{eq:2primetimes2prime}) and Eqs. (\ref{eq:2primeprimetimes2primeprime}). Note that since $2^a = 1^a\x 2$, the reader may prefer to think of the phase $\w$ as coming from the $1'$ and $1''$ reps, thereby choosing the Clebsch-Gordan coefficients of all three doublets to be the same.}
\[(\chi\xi')_3 = \ml -i(\chi_1\xi'_1+\chi_2\xi'_2)\\ -\w^*(\chi_1\xi'_1-\chi_2\xi'_2)\\ +\w\,(\chi_1\xi'_2+\chi_2\xi'_1) \mr\;.\]
Notice the appearance of the phase $\w = e^{\,i2\pi/3}$. This is due to the fact that $2^a = 1^a\x 2$, as we derive explicitly in Eq. (\ref{eq:generatorsof2primeand2primeprime}). Similarly, if we consider the doublet $\xi'' \sim 2''$ then the product $(\chi\xi'') \sim 2\x2'' = 1''\+3$ contains the non-invariant singlet $(\chi\xi'')_{1''} = \chi_1\xi''_2-\chi_2\xi''_1 \sim 1''$ and the triplet
\[(\chi\xi'')_3 = \ml -i(\chi_1\xi''_1+\chi_2\xi''_2)\\ -\w\,(\chi_1\xi''_1-\chi_2\xi''_2)\\ +\w^*(\chi_1\xi''_2+\chi_2\xi''_1) \mr\;.\]
Another group theoretic fact of $T'$ is that $2'\x2' = 2\x2''$ and $2''\x2'' = 2\x2'$, as is made clear by the fact that $2^a = 1^a\x 2$ and $1^a\x 1^b = 1^{a+b}$. So by looking at the above rules we also know how to multiply two $\xi' \sim 2'$-type doublets and two $\xi''\sim 2''$-type doublets. 
\\\\
From studying $T \simeq A_4$, we know how to multiply two singlets and two triplets, and from the above we can multiply two doublets. In $SU(2)$, we also have $2\x3 = 2\+4$, where the 4 of $SU(2)$ is the completely symmetric three-index tensor. Under restriction of $SU(2)$ to the subgroup $T'$, the 4 of $SU(2)$ breaks up into doublets of $T'$ as $4 \to 2'\+2''$, which we show explicitly in Appendix \ref{appendix2}. We therefore find the multiplication rule $2\x3 = 2\+2'\+2''$ in the double tetrahedral group. Let $\chi  = (\chi_1,\chi_2) \sim 2$ of $T'$ as before, and let $\phi = (\phi_1,\phi_2,\phi_3) \sim 3$ be a triplet of $T'$. The explicit construction of the rule $(\chi\phi) \sim 2\x3 = 2\+2'\+2''$ is
\begin{align*}
&(\chi\phi)_2 = \ml -\sqrt2\phi_+\chi_2-i\,\phi_3\chi_1\\ +\sqrt2\phi_-\chi_1+i\,\phi_3\chi_2 \mr\\
&(\chi\phi)_{2'} = \ml (+\phi_+-i2\sqrt3\phi_-)\chi_2-i\frac{1}{\sqrt2}\phi_3\chi_1\\ (-\phi_-+i2\sqrt3\phi_+)\chi_1+i\frac{1}{\sqrt2}\phi_3\chi_2 \mr\\
&(\chi\phi)_{2''} = \ml (+\phi_++i2\sqrt3\phi_-)\chi_2-i\frac{1}{\sqrt2}\phi_3\chi_1\\ (-\phi_--i2\sqrt3\phi_+)\chi_1+i\frac{1}{\sqrt2}\phi_3\chi_2 \mr\\
\end{align*}
where $\phi_\pm \equiv \frac{1}{\sqrt2}(\phi_1\pm i\,\phi_2)$. Note that $(\phi_-)^\da \neq \phi_+$ unless $\phi_1$ and $\phi_2$ are real.
\\\\
Just as multiplying a doublet $\chi \sim 2$ and a triplet $\phi \sim 3$ yields all three inequivalent doublets $2\+2'\+2''$, multiplying a different doublet $\chi' \sim 2'$ with a triplet $\phi$ also yields all three doublets. Write $2^a \x 3 = 2^a\+2^{a+1}\+2^{a+2}$ where the superscripts are defined mod 3 as usual. The leftmost doublet on the right-hand side of the equation is always formed by contracting with an epsilon tensor $\e^{ij}$, so that it maintains the same transformation properties as the doublet $2^a$ on the left-hand side of the equation. The other two doublets $2^{a+1}$ and $2^{a+2}$ come from decomposing the $4$ of $SU(2)$. So to multiply $\chi'' \sim 2''$ with $\phi \sim 3$, cyclically permute the labels $2,2',2''$ in the rules for $2\x3$ shown above. For $\chi'\sim 2'$ with $\phi \sim 3$, anticyclically permute the labels $2,2',2''$. This is analogous to the multiplication rule $1^a\x1^b = 1^{a+b}$ familiar from the tetrahedral group.
\\\\
With the above multiplication rules, we are now ready to propose a theory of neutrino mixing based on the double tetrahedral group.\footnote{A word of caution is in order here since some of the Clebsch-Gordan coefficients are complex. For example, the symmetrized product of two copies of the doublet $\chi \sim 2$ forms a triplet $(\chi\chi)_3 = \left(\,-i(\chi_1^2+\chi_2^2),\, -(\chi_1^2-\chi_2^2),\, 2\chi_1\chi_2 \right)$,
which implies that the conjugate triplet $(\chi^\da\chi^\da)_3 = \left(\, +i(\chi_1^{\da 2}+\chi_2^{\da2}),\, -(\chi_1^{\da2}-\chi_2^{\da2}),\, 2\chi_1^\da\chi_2^\da \right)$ contributes to the neutrino mass matrix.}
%\\\\
%\textbf{IV. The Third Neutrino is Special.}
%\\\\
\section{The Third Neutrino is Special}\label{sec:nu3}
A popular theoretical ansatz for $V$ is the ``tribimaximal mixing matrix" \cite{wolf,hps,hezee}
\[V_{\text{TB}} = \ml -\frac{2}{\sqrt6}&\frac{1}{\sqrt3}&0\\ \frac{1}{\sqrt6}&\frac{1}{\sqrt3}&\frac{1}{\sqrt2}\\ \frac{1}{\sqrt6}&\frac{1}{\sqrt3}&-\frac{1}{\sqrt2} \mr\;.\]
Any orthogonal 3-by-3 matrix can be written as a product of three independent rotations, but since $(V_{\text{TB}})_{e3} = 0$ we can write tribimaximal mixing as a product of only two independent rotations \cite{zeeangles}:
\begin{equation}
V_{\text{TB}} = \ml 1&0&0\\0&\frac{1}{\sqrt2}&\frac{1}{\sqrt2}\\ 0&\frac{1}{\sqrt2}&-\,\frac{1}{\sqrt2} \mr\ml -\sqrt{\frac{2}{3}}&\frac{1}{\sqrt3}&0\\ \frac{1}{\sqrt3}&\sqrt{\frac{2}{3}}&0\\0&0&1 \mr\;.
\end{equation}
On purely phenomenological grounds, we know that this mixing matrix is at least approximately correct in the sense that $|V_{e3}|$ is known to be small.
\\\\
Theoretically, the possibility that $V$ can be decomposed into only two independent rotations may provide a hint for some underlying structure in the lepton sector. The definition $V = U_L^{-1}U_\nu$ suggests that we look for a theory in which
\[U_L^{-1} = U_L = \ml 1&0&0\\0&\frac{1}{\sqrt2}&\frac{1}{\sqrt2}\\0&\frac{1}{\sqrt2}&-\frac{1}{\sqrt2} \mr\;\;\;\text{and}\;\;\;\; U_\nu = U_\nu^{-1} = \ml -\sqrt{\frac{2}{3}}&\frac{1}{\sqrt3}&0\\ \frac{1}{\sqrt3}&\sqrt{\frac{2}{3}}&0\\0&0&1 \mr\;.\]
The form for $U_L$ suggests we treat the second and third families as a doublet and the first family as a singlet, while the form for $U_\nu$ suggests we treat the first and second families as a doublet and the third family as a singlet. Since the charged lepton mass matrix is made from two types of fields, $\{e_i\}_{i\,=\,1}^3$ and $\{e_i^c\}_{i\,=\,1}^3$, while the neutrino mass matrix is made from only one type of field $\{\nu_i\}_{i\,=\,1}^3$, we take the suggestion from the neutrino mass sector seriously and use the extra freedom in the charged lepton sector to adjust the mixing matrix as needed. We thus choose the transformation properties
\begin{equation}\label{eq:nureps}
\nu \equiv \ml \nu_1\\ \nu_2 \mr \sim 2\;,\;\; \nu_3 \sim 1\;\;\;\text{under}\;\;T'
\end{equation}
for the neutrinos. The neutrino mass matrix $M_\nu$ is made from the singlet operator\footnote{The other possible singlet $(\nu\nu)_1 = \e^{ij}\nu_i\nu_j$ is zero since $\nu_i\nu_j \equiv \e_{\al\beta}\nu_i^\al \nu_j^\beta$ is symmetric in $i$ and $j$. See footnote $2$.} $\op_1 \equiv \nu_3\nu_3 \sim 1$, the doublet operator $\op_2 \equiv \ml \nu_1\\ \nu_2 \mr \nu_3 \sim 2$, and the triplet operator
\[\op_3 \equiv (\nu\nu)_3 = \ml -i(\nu_1\nu_1+\nu_2\nu_2)\\ -(\nu_1\nu_1-\nu_2\nu_2)\\ \nu_1\nu_2+\nu_2\nu_1 \mr \sim 3\;.\]
Let $\phi$ be an electroweak Higgs doublet that transforms as a triplet under $T'$. Since $3\x3 = 1\+1'\+1''\+3_1\+3_2$, we can form the singlet\footnote{We can also form the non-invariant singlets $(\phi\phi)_{1'} = \phi_1^2+\w\phi_2^2+\w^2\phi_3^2$ and $(\phi\phi)_{1''} = \phi_1^2+\w^2\phi_2^2+\w\phi_3^2$, but these will not form $T'$-invariant products with the neutrinos.} $(\phi\phi)_1 = \sum_{i\,=\,1}^3\phi_i^2$ and the triplet\footnote{The two triplets $3_1$ and $3_2$ are equal in this case. In other words, since $\phi_i\phi_j$ is symmetric in $i$ and $j$, the antisymmetric combination is zero.} $(\phi\phi)_3 = (\phi_2\phi_3,\phi_3\phi_1,\phi_1\phi_2)$. The singlet couples to $\op_1$, and the triplet couples to $\op_3$:
\[[(\phi^\da\phi^\da)_3\op_3]_1 = -(\phi_1^\da+i\phi_2^\da)\phi_3^\da\nu_1\nu_1+(\phi_1^\da-i\phi_2^\da)\phi_3^\da\nu_2\nu_2+\phi_1^\da\phi_2^\da(\nu_1\nu_2+\nu_2\nu_1)\;.\]
Since we have not introduced a doublet scalar field, nothing couples to $\op_2$. The neutrino mass matrix comes from the Lagrangian\footnote{The factor of $\frac{1}{3}$ in front of the coupling $z_2$ is just for the aesthetic convenience of canceling the factor 3 that will result from the chosen vacuum alignment.} $\la = -\,\frac{1}{\Ld}z_1[(\phi^\da\phi^\da)_3\op_3]_1-\,\frac{1}{3\Ld}z_2(\phi^\da\phi^\da)_1\op_1+h.c.$ .
\\\\
We want the upper-left block of the matrix $U_\nu$ to have off-diagonal terms, so take $\phi_1 = \phi_2 = \phi_3 \equiv \phi \neq 0$. Then the neutrino mass matrix is
\begin{equation}\label{eq:Mnu}
M_\nu = m_\nu \ml -\sqrt2\,e^{+i\pi/4}&1&0\\1&+\sqrt2\,e^{-i\pi/4}&0\\0&0&z \mr
\end{equation}
where $m_\nu \equiv \frac{2}{\Ld}z_1\phi^{\da 2}$ and $z \equiv z_2/z_1$. Notice the factors of $\sqrt2$ and the complex phases $e^{\pm i\pi/4}$ forced upon us by group theory. The 2-by-2 complex matrix\footnote{The parameters $\al$ and $\beta$ are real.}
\[M = \ml -\al\,e^{+i\beta}&1\\1&\al\,e^{-i\beta} \mr\]
is put into the form $U_\nu^TM_\nu U_\nu = \text{diag}(m,m)$ with $m = \sqrt{\al^2+1}$ by the unitary matrix
\[U_\nu = \ml e^{-i\beta/2}&0\\0&e^{+i\beta/2} \mr \ml -\cos\ta&\sin\ta\\ \sin\ta&\cos\ta \mr\ml i&0\\0&1 \mr \]
where $\tan(2\ta) \equiv 1/\al$. We thus satisfy the requirement $m_2^2-m_1^2 \ll |m_3^2-m_1^2|$ to lowest order, but we will need to modify the mass matrix to split the degeneracy $m_1 = m_2$. 
Numerically we have $\al = \sqrt2$ and $\beta = \pi/4$, so we predict the unitary matrix
\begin{equation}\label{eq:Unuwrong}
U_\nu \approx \ml e^{-i\pi/8}&0&0\\0&e^{+i\pi/8}&0\\0&0&1 \mr\ml -0.95&0.30&0\\ 0.30&0.95&0\\ 0&0&1 \mr\ml i&0&0\\0&1&0\\0&0&1 \mr
\end{equation}
which is a rotation through $\ta \approx 0.31 \sim 18^\circ$ in the $(1,2)$-plane. Recall that we wanted
\[U_\nu = \ml -\sqrt{\frac{2}{3}}&\frac{1}{\sqrt3}&0\\ \frac{1}{\sqrt3}&\sqrt{\frac{2}{3}}&0\\0&0&1 \mr \approx \ml -0.82&0.58&0\\0.58&0.82&0\\0&0&1 \mr \]
which is a rotation through $\sim 35^\circ$ in the $(1,2)$-plane. Since we need an extra $\sim 17^\circ$, the modification we need is not a small perturbation.
\\\\
At this point we emphasize that our use of tribimaximal mixing is meant only to motivate the factorization of the mixing matrix into a rotation in two stages. In our model, the neutrino mass matrix will essentially determine the upper left 2-by-2 block of the mixing matrix, but $U_\nu$ will not necessarily be a rotation purely in one plane. This reflects the fact that although tribimaximal mixing is compatible with data, it is certainly not the only option. The idea is simply that the lower right 2-by-2 block of $V$ will be adjusted by the charged lepton sector.
%\\\\
%\textbf{V. Charged Lepton Sector.}
%\\\\
\section{Charged Lepton Sector}\label{sec:charged}
Before trying to remedy the problems with $M_\nu$, consider the charged leptons. From our choices of transformation properties in the neutrino sector [Eq.(\ref{eq:nureps})], we inherit the assignments
\begin{equation}
\ml e_1\\ e_2 \mr \sim 2\;,\;\; e_3 \sim 1\;.
\end{equation}
To generate a charged lepton mass matrix that can be diagonalized by something of the schematic form
\begin{equation}\label{eq:UL}
 U_L \sim U_R \sim \ml 1&0&0\\0&\times&\times\\ 0&\times&\times \mr
\end{equation}
we treat the conjugate leptons $e_i^c$ as singlets of $T'$. Just as in $T \simeq A_4$, we have the choice of three singlets $1,1'$ and $1''$. Even though here we are just using low-energy effective field theory, we can imagine that the discrete symmetry $T'$ should arise from breaking an $SU(2)$ symmetry at a higher energy scale. The $1'$ and $1''$ would arise from the decomposition $5 \to 1'\+1''\+3$ when $SU(2)$ is restricted to the subgroup $T'$. 
\\\\
Meanwhile, in the charged lepton mass matrix $M_\ell$ we want the terms $(M_\ell)_{32}$ and $(M_\ell)_{33}$ to be nonzero, while we want $(M_\ell)_{31} = 0$. We thus pick the assignments $e_2^c \sim 1'$ and $e_3^c \sim 1''$ and introduce the Higgs fields $\ph' \sim 1'$ and $\ph'' \sim 1''$ to couple to them, and we pick $e_1^c \sim 1$ but do not introduce a singlet $\ph \sim 1$ that would be invariant under $T'$. To summarize, we choose
\begin{equation}
e_1^c \sim 1\;,\;\; e_2^c \sim 1'\;,\;\; e_3^c \sim 1''\;.
\end{equation}
The charged lepton mass matrix therefore comes from the three doublet operators
\[\op_1 \equiv \ml e_1\\ e_2 \mr e_1^c \sim 2,\,\op_2 \equiv \ml e_1\\ e_2 \mr e_2^c\sim 2',\,\op_3 \equiv \ml e_1\\ e_2 \mr e_3^c \sim 2''\]
and the three singlet operators
\[\op_4 \equiv e_3e_2^c\sim 1',\, \op_5 \equiv e_3e_3^c \sim 1'',\, \op_6 \equiv e_3e_1^c \sim 1.\]
The Lagrangian $\la = -y_4\ph''\op_4-y_5\ph'\op_5+h.c.$ implies the form $M_\ell \sim \ml 0&0&0\\0&0&0\\0&\times&\times \mr$, which so far only gives mass to the tau. To give masses to the electron and muon, introduce new Higgs fields $\xi,\,\xi'$ and $\xi''$ that transform as $2,\,2'$ and $2''$, respectively, under $T'$. First, $\xi \sim 2$ implies $(\xi\op_1)_1 = (\xi_1e_2-\xi_2e_1)e_1^c = \xi_1\,e_2e_1^c-\xi_2\,e_1e_1^c$. Since we want $(M_\ell)_{21} = 0$ while $(M_\ell)_{11} \neq 0$, we want $\xi_1 = 0$ and $\xi_2 \equiv \xi \neq 0$. Since $\xi' \sim 2'$ and $\op_3 \sim 2''$, the property $2'\x 2'' = 1\+...$ implies we can form the invariant $(\xi'\op_3)_1 = (\xi_1'e_2-\xi_2'e_1)e_3^c = \xi_1'\,e_2e_3^c-\xi_2'\,e_1e_3^c$. We want $(M_\ell)_{23} \neq 0$ while $(M_\ell)_{13} = 0$, so we want $\xi_1' \equiv \xi' \neq 0$ and $\xi_2' = 0$. Similarly, we can form the invariant $(\xi''\op_2)_1 = \xi_1''\,e_2e_2^c-\xi_2''\,e_1e_2^c$. We want $(M_\ell)_{22} \neq 0$ while $(M_\ell)_{12} = 0$, so we want $\xi''_1 \equiv \xi'' \neq 0$ and $\xi_2 = 0$. With these vacuum alignments, the Lagrangian
\[\la = -y_1(\xi\op_1)_1-y_2(\xi''\op_2)_1-y_3(\xi'\op_3)_1-y_4\ph''\op_4-y_5\ph'\op_5+h.c.\]
implies the charged lepton mass matrix
\begin{equation}
M_\ell = \ml -y_1\xi&0&0\\ 0&y_2\xi''&y_3\xi'\\ 0&y_4\ph''&y_5\ph' \mr\;.
\end{equation}
Thus we have 5 independent entries to adjust, which arise from 5 different Higgs fields: $\xi \sim 2,\,\xi' \sim 2',\,\xi'' \sim 2'',\,\ph' \sim 1'$ and $\ph'' \sim 1''$. The point is that we have achieved the desired zeros, so that $M_\ell$ can be diagonalized by matrices of the form in Eq. (\ref{eq:UL}).
%\[U_L \sim U_R \sim \ml 1&0&0\\0&\times&\times\\ 0&\times&\times \mr\;.\]
This is of course too many free parameters to claim any true understanding of the charged lepton sector, but our purpose here is to illuminate some structure in the neutrino sector without worrying too much about the charged leptons. 
%\\\\
%\textbf{VI. A Mixing Matrix Consistent with Data.}
%\\\\
\section{A Mixing Matrix Consistent with Data}\label{sec:compatible}
From Section \ref{sec:nu3} with only the $T'$-triplet Higgs $\phi \sim 3$, we had the Lagrangian
\[\la_{\text{old}} = -\,\frac{1}{\Ld}z_1[(\phi^\da\phi^\da)_3\op_3]_1-\,\frac{1}{3\Ld}z_2(\phi^\da\phi^\da)_1\op_1+h.c.\]
As mentioned, the immediate problem with this model as it stands is that the top row of the mixing matrix $V = U_L^{-1}U_\nu$ is wrong: $|V_{e1}|$ can be at most 0.86 and $|V_{e2}|$ must be at least 0.50, whereas as mentioned earlier this model predicts $|V_{e1}| \approx 0.95$ and $|V_{e2}| \approx 0.30$. Since we already knew that we need to modify the neutrino mass matrix to get $m_2^2-m_1^2 > 0$, we now remedy both deficiencies by returning to the neutrino sector.
\\\\
We can break the degeneracy $m_1 = m_2$ by including a scalar doublet\footnote{We remind the reader that, as stated at the end of Section \ref{sec:minimal}, we assume that the Higgs fields in the charged lepton sector do not contribute to the neutrino mass matrix.} $\chi \sim 2$ to couple to $\op_2 = \ml \nu_1\\ \nu_2 \mr\nu_3 \sim 2$ through the group theoretic multiplication rule $2\x3 = 2\+2'\+2''$. Explicitly, we have 
\[(\chi^\da\phi^\da)_2 = \ml -\sqrt2\chi_2^\da\phi_+^\da+i\chi_1^\da\phi_3^\da\\ +\sqrt2\chi_1^\da\phi_-^\da-i\chi_2^\da\phi_3^\da \mr \;.\]
This can couple to $\op_2$ to form the $T'$-invariant term $[(\chi^\da\phi^\da)_2\op_2]_1 = (-\sqrt2\,e^{-i\pi/4}\chi_2^\da+i\chi_1^\da)\phi^\da\nu_2\nu_3 +(-\sqrt2\,e^{+i\pi/4}\chi_1^\da+i\chi_2^\da)\phi^\da\nu_1\nu_3$, where we have again chosen the vacuum alignment $\phi_1 = \phi_2 = \phi_3 \equiv \phi$.
\\\\
With a doublet $\chi$, we also have a new triplet from $2\x2 = 1\+3$:
\[(\chi^\da\chi^\da)_3 = \ml +i(\chi_1^{\da2}+\chi_2^{\da2})\\ -(\chi_1^{\da2}-\chi_2^{\da2})\\ 2\chi_1^\da\chi_2^\da \mr\;.\]
This couples to the triplet $\op_3 = (\nu\nu)_3 = \left( -i(\nu_1\nu_1+\nu_2\nu_2),-(\nu_1\nu_1-\nu_2\nu_2),\nu_1\nu_2+\nu_2\nu_1\right)$. To the previous neutrino mass terms, we add the Lagrangian\footnote{Again we introduce factors of $2$ and $\frac{1}{2}$ for later aesthetic convenience.}
\[\la_{\text{new}} = -\,\frac{2}{\Ld}z_1'[(\chi^\da\phi^\da)_2\op_2]_1-\,\frac{1}{2\Ld}z_2'[(\chi^\da\chi^\da)_3\op_3]_1+h.c.\;.\]
Note that we do not gain a new singlet since $\e^{ij}\chi_i\chi_j = 0$. 
\\\\
We can make $m_1 \neq m_2$ by changing the value of $(M_\nu)_{22}$ while leaving alone $(M_\nu)_{11}$ and $(M_\nu)_{12} = (M_\nu)_{21}$. Take $\chi_1 = 0$ and $\chi_2 \equiv \chi \neq 0$, so that the neutrino mass matrix is
\[M_\nu = \frac{\phi^{\da2}}{\Ld}\ml -z_1\sqrt2\,e^{+i\pi/4}&z_1&i\,rz_1' \\z_1&z_1\sqrt2\,e^{-i\pi/4}+r^2z_2'&-r\sqrt2\,e^{-i\pi/4}z_1'\\i\,rz_1' &-r\sqrt2\,e^{-i\pi/4}z_1'&z_2 \mr\;,\]
where $r \equiv \phi^{\da}/\chi^{\da}$. Two remarks are in order: First, we are forced to introduce rotations in both the (1,3)-plane and the (2,3)-plane.\footnote{There are two exceptions to this: the specific vacuum alignment $\chi_2^\da/\chi_1^\da = \frac{1}{\sqrt2}\,i\,e^{+i\pi/4}$ sets the $\nu_2\nu_3$ term to zero, and the specific alignment $\chi_2^\da/\chi_1^\da = -\sqrt2\,i\,e^{+i\pi/4}$ sets the $\nu_1\nu_3$ term to zero.} Second, we are forced to introduce complex phases through the factors of $e^{-i\pi/4}$ and $i = e^{+i\pi/2}$.
\\\\
To understand the effects of the new terms, rewrite the above matrix as
\begin{equation}\label{eq:Mnufixed}
M_\nu = m_\nu \ml -\sqrt2\,e^{+i\pi/4}&1&i\,\e \\ 1&+\sqrt2\,e^{-i\pi/4}+\del&-\sqrt2\,e^{-i\pi/4}\e\\ i\,\e&-\sqrt2\,e^{-i\pi/4}\e&z \mr
\end{equation}
where $m_\nu \equiv \frac{2}{\Ld}z_1\phi^{\da2}$, $\e \equiv \frac{z_1'}{z_1}r$, $\del \equiv \frac{z_2'}{z_1}r^2$ and $z \equiv \frac{z_2}{z_1}$ (and $r \equiv \phi^\da/\chi^\da$ as before). When $\e = \del = 0$, we recover Eq. (\ref{eq:Mnu}). The goal is to modify the first row of $U_\nu$ in Eq. (\ref{eq:Unuwrong}) significantly without drastically altering the other rows. As an example, for $\e = 6,\,\del = -4,\, z = 21$ we get $R \approx 24.4$, which is in the allowed range, and 
\begin{equation}
U_\nu \approx \ml e^{-i0.30}\!\!&0&0\\0&e^{-i2.25}\!\!&0\\0&0&-1 \mr \ml -0.82&0.54&0.22\,e^{-i1.33}\\0.54\,e^{+i0.11}&0.78\,e^{-i0.05}&0.32\\0.22\,e^{-i0.88}&0.32\,e^{+i0.35}&-0.92 \mr \ml e^{-i2.84}\!\!&0&0\\0&e^{+i2.30}\!\!&0\\0&0&1 \mr\;.
\end{equation}
\\
We have chosen\footnote{We remind the reader that we are not explicitly trying to reproduce tribimaximal mixing. Instead, we are trying to maximize experimental interest in measuring $V_{e3}$ and thus in constraining this model. See the comment at the end of Section~\ref{sec:nu3}.} the values of $\e,\,\del$ and $z$ to maximize the value of $|(U_\nu)_{13}|$, which is at the upper limit of $0.22$ for $|V_{e3}|$. Since the first row and first column fit data, we want a rotation in the (2,3)-plane to increase $|(U_\nu)_{23}|$ and $|(U_\nu)_{32}|$ and to decrease $|(U_\nu)_{22}|$ and $|(U_\nu)_{33}|$. If we keep $\e = 6,\,\del = -4$ and $z = 21$, then a rotation
\begin{equation}
U_L = \ml 1&0&0\\0&-\cos\ta&\sin\ta\\ 0&\sin\ta&\cos\ta \mr\;\;\text{ with }\;\;\ta = 0.4 \sim 23^\circ
\end{equation}
and the definition $V \equiv U_L^{-1}U_\nu$ gives
\[|V| \approx \ml 0.82&0.54&0.22\\ 0.52&0.61&0.59\\ 0.24&0.58&0.78 \mr\]
which is compatible with oscillation data. 
%\\\\
%\textbf{VII. CP Violation.}
%\\\\
%\pagebreak
\section{CP Violation}\label{sec:CP}
In this model $|V_{e3}|$ can be at its empirical upper bound, so we should ask whether CP is conserved. This means we need to put the mixing matrix into the form\footnote{Nonzero phases in $\M \equiv \text{diag}(e^{\,i\rho},e^{\,i\sigma},1)$ also indicate CP violation, but here we are concerned with oscillation experiments, which cannot observe $\M$. The phases in $\mathcal K \equiv \text{diag}(e^{\,i\ka_1},e^{\,i\ka_2},e^{\,i\ka_3})$ are unphysical. See Eq. (\ref{eq:KPM}) for the notation.} $V = \mathcal K V_{\text{PMNS}} \M$ and find the value of $\del_{\text{CP}} \equiv -\arg\,(V_{\text{PMNS}})_{e3}$. For the particular values of the parameters quoted previously, this decomposition reads
\begin{equation}
V \approx \ml e^{-i0.92}\!\!&0&0\\0&e^{+i0.40}&0\\0&0&\!\!e^{+i3.02} \mr\ml -0.82&0.54&0.22\,e^{-i0.72}\\ 0.52\,e^{+i0.14}&0.61\,e^{-i0.08}&0.59\\ 0.24\,e^{-i0.40}&0.58\,e^{+i0.11}&-0.78 \mr \ml e^{-i2.22}\!\!&0&0\\0&e^{+i2.92}&0\\0&0&1 \mr\;.
\end{equation}
\\
The CP violating angle in the PMNS matrix is $\del_{\text{CP}} \approx 0.72$. Thus this model violates CP even though we have assumed all of the coupling constants and Higgs vacuum expectation values to be real. The choice of $T'$ as the flavor group can result in a geometric source of CP violation, meaning that physically observable complex phases result purely from group theory.
%\pagebreak\\
%\textbf{VIII. Discussion.}
%\\\\
\section{Discussion}\label{sec:discussion}
We have constructed a model for neutrino mixing based on the double tetrahedral group, $T'$, under which the first two neutrinos transform as a doublet and the third neutrino is a singlet. Introducing Higgs fields $\phi \sim 3$ and $\chi \sim 2$ of $T'$, the neutrino mass matrix in the flavor basis requires an additional rotation through $\sim 23^\circ$ in the $(2,3)$-plane to fit oscillation data, which we accommodate using the charged leptons. Perhaps a more elaborate framework in the charged lepton sector could ``predict" this extra rotation with fewer free parameters, since as it stands the angle from the charged lepton sector is accommodated simply by tuning the relevant parameters.
\\\\
The most important feature of our model is that it violates $CP$ even when all coupling constants and Higgs vacuum expectation values are real. This is as in the $SU(5)\x T'$ model of Chen and Mahanthappa \cite{chen}. Since $|V_{e3}|$ can be at the upper limit of $0.22$ in this model, we predict that neutrino oscillations violate $CP$. This is where our model differs phenomenologically from that of \cite{chen}, since our $V_{e3}$ can be large.
\\\\
As a final remark on this model, if we insist that $T'$ symmetry is a remnant of a high energy $SU(2)$ symmetry, then the fields $e_2^c \sim 1'$ and $e_3^c \sim 1''$ should come from a 5 of $SU(2)$, which as noted earlier breaks up as $5 \to 1'\+1''\+3$. This suggests a new as of yet unobserved triplet fermion, which presumably has a mass comparable to the scale of breaking $SU(2)$ to $T'$. We then have to worry about canceling anomalies, but in the absence of a larger framework we will not pursue this idea further.
%\\\\
%\textbf{Appendix: Group Theory.}
%\\\\
\appendix
\section{Group Theory}
For the convenience of the reader, we review the group theory required to understand the tetrahedral and double tetrahedral groups and derive explicitly all relevant results.
%\\\\
%\textbf{A.1: $SO(3)$ and the Tetrahedral Group, $T$.}
%\\\\
\subsection{$SO(3)$ and the Tetrahedral Group, $T$}
Any rotation in three dimensions can be parametrized as $R(\hat n, \ta) = n_in_j + (\delta_{ij}-n_in_j)C - \e_{ijk}n_kS$, where $C \equiv \cos\ta$, $S \equiv \sin\ta$, and $\sum_{i\,=\,1}^3n_i^2 \equiv 1$. The collection of all such rotations is the 3-dimensional representation of $SO(3)$.
\\\\
The particular collection of rotations that also leaves invariant a regular tetrahedron is called the tetrahedral group, $T$. Choose coordinates for which the vertices of the regular tetrahedron with sides of length $\sqrt{8/3}$ lie along the axes
\[\hat 1 \equiv \frac{1}{\sqrt3}(1,1,1),\;\hat 2 \equiv \frac{1}{\sqrt3}(-1,-1,1),\;\hat 3 \equiv \frac{1}{\sqrt3}(-1,1,-1),\;\hat 4 \equiv \frac{1}{\sqrt3}(1,-1,-1)\;.\]
The complete collection of symmetries of the tetrahedron is
\[I,\,\{r_1,r_2,r_3\},\{c,r_1cr_1,r_2cr_2,r_3cr_3\},\,\{a,r_1ar_1,r_2ar_2,r_3ar_3\}\]
which makes a total of 12 elements. The notation is $r_1 \equiv R(\pi,\hat x) = \text{diag}(1,-1,-1),\,r_2 \equiv R(\pi,\hat y) = \text{diag}(-1,1,-1),\,r_3 \equiv R(\pi,\hat z) = \text{diag}(-1,-1,1),\, c \equiv R(2\pi/3,\hat 1),\,a \equiv R(-2\pi/3,\hat 1)$. The matrices $c$ and $a$ implement cyclic and anticyclic permutations, respectively. The braces ``$\{...\}$" separate the elements into equivalence classes under the operation of conjugation, which are called conjugacy classes.
\\\\
The explicit construction above is the 3-dimensional representation of $T$ inherited from the continuous rotation group $SO(3)$. Also from $SO(3)$, we inherit the invariant 1-dimensional representation, the trace. \\\\
Traceless symmetric tensors, which comprise the 5-dimensional representation of $SO(3)$, fall apart into smaller irreducible representations under restriction to the tetrahedral subgroup.\footnote{Let $n_G$ be the number of elements in a group $G$, and let $n_C$ be the number of conjugacy classes in the group. Let $d_i$ be the dimension of the $i^{\text{th}}$ irreducible representation of the group. There is a theorem that says
\[\sum_{i\,=\,1}^{n_C}d_i^{\,2} = n_G\]
which, for $n_C = 4$ and $n_G = 12$, has the unique solution $d_1 = d_2 = d_3 = 1,\, d_4 = 3$.} Let $M_{ij} \equiv \frac{1}{2}(M_{ij}+M_{ji})-\frac{1}{3}\del_{ij}\,\tr M$ transform as a 5 of $SO(3)$. By explicitly acting on $M_{ij}$ with the elements of $T$, we find that the combination $(M_{23},M_{31},M_{12})$ transforms as a vector, while $M_{11}+\w\, M_{22}+\w^* M_{33}$ and $M_{11}+\w^*M_{22}+\w\, M_{33}$ transform as singlets, which we call $1'$ and $1''$ respectively. ($\w \equiv e^{\,i2\pi/3}$.) In other words, we find $5 \to 1'\+1''\+3$ when $SO(3)$ is restricted to the subgroup $T$.
\\\\
Like the singlet $1$ formed from the trace, the singlets $1'$ and $1''$ are invariant under the rotations by $\pi$. But unlike the singlet $1$, which is invariant under all rotations, the singlets $1'$ and $1''$ are not invariant under the rotations by $\pm 2\pi/3$. To summarize: for two triplets $v \sim 3$ and $w \sim 3$, we have in the tetrahedral group the rule $3\x3 = 1\+1'\+1''\+3_A\+3_S$, where
\begin{align}\label{eq:TCG}
&(vw)_1 = v_1w_1+v_2w_2+v_3w_3 \nonumber \\
&(vw)_{1'} = v_1w_1+\w\,v_2w_2+\w^*v_3w_3\nonumber \\
&(vw)_{1''} = v_1w_1+\w^*v_2w_2+\w\,v_3w_3\nonumber \\
&(vw)_{3_A} = \ml v_2w_3-v_3w_2\\v_3w_1-v_1w_3\\v_1w_2-v_2w_1 \mr\nonumber \\
&(vw)_{3_S} = \ml v_2w_3+v_3w_2\\v_3w_1+v_1w_3\\v_1w_2+v_2w_1 \mr\;.
\end{align}
If desired, we may repackage the symmetric and antisymmetric triplets into the triplets $(vw)_{3_1} = (v_2w_3,v_3w_1,v_1w_2)$ and $(vw)_{3_2} = (v_3w_1,v_1w_3,v_2w_1)$.
\\\\
These multiplication rules constitute everything one needs to know in order to build Lagrangians with tetrahedral symmetry.
%\\
%\textbf{A.2: $SU(2)$ and the Double Tetrahedral Group, $T'$.}
%\\\\
\subsection{$SU(2)$ and the Double Tetrahedral Group, $T'$}\label{appendix2}
The 3-by-3 rotation matrix $R(\hat n,\ta)$ from $SO(3)$ is, to first order in the angle $\ta$
\[R(\hat n,\ta) = I + \ta \ml 0&-n_z&n_y\\ n_z&0&-n_x\\ -n_y&n_x&0 \mr + O(\ta^2) = I+\ta\sum_{a\,=\,1}^3n_a T^a+O(\ta^2)\]
where
\[T^1 \equiv \ml 0&0&0\\0&0&-1\\0&1&0 \mr\;,\;\; T^2 \equiv \ml 0&0&1\\0&0&0\\-1&0&0 \mr\;,\;\; T^3 \equiv \ml 0&-1&0\\1&0&0\\0&0&0 \mr\;.\]
These matrices satisfy the relations $[T^a,T^b] = \sum_{c\,=\,1}^3\e^{abc} T^c$, where $\e^{123} \equiv +1$.
\\\\
The 2-by-2 matrices $t^a \equiv -i\sigma^a/2$, where $\sigma^1 \equiv \ml 0&1\\1&0 \mr,\;\sigma^2 \equiv \ml 0&-i\\i&0 \mr,\;\sigma^3 \equiv \ml 1&0\\0&-1 \mr$ are the Pauli matrices, satisfy the commutation relations $[t^a,t^b] = \e^{abc}t^c$, which are the same as those satisfied by $\{T^a\}_{a\,=\,1}^3$. When we exponentiate the matrices $t^a$, we get
\[e^{-\ta n_at^a} = e^{+i\ta n_a\sigma^a/2} = Ic + i\,n_a \sigma^a s = \ml c+i\,n_zs&+i(n_x-i\,n_y)s\\ +i(n_x+i\,n_y)s&c-i\,n_zs \mr\]
where $c \equiv \cos(\ta/2)$ and $s \equiv \sin(\ta/2)$. This is an arbitrary 2-by-2 unitary matrix with determinant 1, or in other words it is an arbitrary element of the group $SU(2)$. Thus the groups $SU(2)$ and $SO(3)$ are locally isomorphic.
\\\\
It is worth noting that $SO(3)$ is not a subgroup of $SU(2)$. For example, consider rotations purely in the $\hat z$-direction, $R_2(\hat z,\ta) = e^{\,i\ta \sigma^3/2}$. If $SO(3)$ were a subgroup, then the angular restriction $0 < \ta \leq 2\pi$ would form a group. But as soon as one reaches $\ta = 2\pi$ in $SU(2)$, one reaches minus the identity element, $-I$, which is not infinitesimally close to $+I$. In other words, $e^{\,i\ta \sigma^3/2}$ with $0 < \ta \leq 2\pi$ does not form a group, since $(e^{\,i\ta\sigma^3/2})(-I)$ would not be in the group.
\\\\
Since $SO(3)$ is locally isomorphic to $SU(2)$, we might ask if we can define a group based on the collection of operations that leaves a regular tetrahedron invariant, but have the rotation matrices valued in $SU(2)$ rather than in $SO(3)$. Let $R_3(\hat n,\ta)$ be an element of $SO(3)$ and $R_2(\hat n,\ta)$ be an element of $SU(2)$. Although the two groups are locally isomorphic, $R_3(\hat n,2\pi) = +I$ while $R_2(\hat n,2\pi) = -I$. We need to go around the $SO(3)$ group space twice in order to make a full lap around the $SU(2)$ group space, and this is what is meant by $SU(2)$ being the double cover of $SO(3)$. 
\\\\
As discussed, the tetrahedral group $T$ contains 12 elements that fall into 4 conjugacy classes:
\[I,\,\{r_1,r_2,r_3\},\{c,r_1cr_1,r_2cr_2,r_3cr_3\},\,\{a,r_1ar_1,r_2ar_2,r_3ar_3\}\;.\]
Since $T'$ double covers $T$, we double the number of elements to 24:
\begin{align*}
&\;\;\;I,\,r_1,r_2,r_3,\,c,r_1cr_1,r_2cr_2,r_3cr_3,\,a,r_1ar_1,r_2ar_2,r_3ar_3\\
&-I,\,-r_1,-r_2,-r_3,\,-c,-r_1cr_1,-r_2cr_2,-r_3cr_3,\,-a,-r_1ar_1,-r_2ar_2,-r_3ar_3\;.
\end{align*}
Naively we might expect the number of conjugacy classes to double from 4 to 8, but there is a subtlety. Using the explicit form of $R_2(\hat n,\ta)$, we see that $r_i = i\sigma_i$. This implies, for example,
\[r_1r_3r_1 = -r_3\]
rather than $+r_3$, so that $r_i$ and $-r_i$ are actually in the same conjugacy class. Therefore there are 7, not 8, conjugacy classes in total.\footnote{The theorem
\[\sum_{i\,=\,1}^7d_i^2 = 24\]
along with the $T\simeq A_4$ representations $d_1 = d_2 = d_3 = 1,\,d_4 = 3$ then implies $d_5 = d_6 = d_7 = 2$. The 1 is the invariant trace inherited from $SO(3)\simeq SU(2)$, and the 2 is the doublet inherited from the defining representation of $SU(2)$. Just as for $T$ we found two additional singlet irreducible representations $1'$ and $1''$, here for $T'$ we deduce the existence of two additional doublet representations $2'$ and $2''$. In $T$, the singlets $1'$ and $1''$ are invariant under the rotations $r_i$, but they transform under the cyclic and anticyclic permutations. Similarly, if $c$ generates cyclic permutations on the $2$, then $c' \equiv \w\,c$ generates cyclic permutations on the $2'$, and $c'' \equiv \w^*c$ generates cyclic permutations on the $2''$. Also, $a' \equiv \w^*a$ generates anticyclic permutations on the $2'$, and $a'' \equiv \w\,a$ generates anticyclic permutations on the $2''$. This is all consistent with the explicit calculations in the rest of this section.}
\\\\
The finite group $T'$ is generated by repeated multiplication of elements. We can use
\begin{equation}\label{eq:Tprimegenerators}
r \equiv R_2(\hat z,\pi) = \ml i&0\\0&-i \mr = i\sigma_z\;\;\text{and}\;\; c \equiv R_2(\hat 1,\frac{2\pi}{3}) = \frac{\tau}{\sqrt2}\ml 1&1\\i&-i \mr
\end{equation}
where $\tau \equiv e^{\,i\pi/4}$. Note that $r^2 = -I$ and $c^3 = -I$, so that $r^4 = c^6 = I$, as expected. Just as we construct explicitly the $1'$ and $1''$ irreps by ``discovering" the decomposition $5 \to 1'\+1''\+3$ when the 5 of $SO(3)$ is restricted to the tetrahedral subgroup, we will now discover the decomposition $4 \to 2'\+2''$ when the 4 of $SU(2)$ is restricted to the double tetrahedral subgroup.
\\\\
Let $i = 1,2$ be an index for the defining representation of $SU(2)$. The two-index tensor $M_{ij}$ transforms reducibly under $SU(2)$ as $1_A\+3_S$. Under $c$, the components of the symmetric tensor transform as
\begin{align*}
&-(M_{11}-M_{22})\,\, \to -i(M_{11}+M_{22})\\
&-i(M_{11}+M_{22}) \to 2M_{12}\\
&\;\;\qquad\qquad2M_{12}\, \to -(M_{11}-M_{22}),
\end{align*}
and under $r$ they transform as $M_{11} \to -M_{11},\,M_{22} \to -M_{22}$ and $M_{12} \to +M_{12}$. A triplet $(M_1,M_2,M_3) \sim 3$ of $SU(2)$ transforms under $c$ as $(M_1,M_2,M_3) \to (M_3,M_1,M_2)$ and under $r$ as $(M_1,M_2,M_3) \to (-M_1,-M_2,+M_3)$, so we can repackage the symmetric tensor as a vector:
\begin{equation}\label{eq:2times2}
\ml M_1\\ M_2 \\ M_3 \mr = \ml -i(M_{11}+M_{22})\\-(M_{11}-M_{22})\\ 2M_{12} \mr\;.
\end{equation}
Inverting these relations gives $M_{11} = \frac{i}{2}(M_1+iM_2),\,M_{12} = \frac{1}{2}M_3$ and $M_{22} = \frac{i}{2}(M_1-iM_2)$.
\\\\
Now consider the 3-index tensor $N_{i(jk)}$, where the notation means that only the last two indices are symmetrized. This tensor transforms as the $2\x3$-representation of $SU(2)$. This representation is reducible, since we can symmetrize and antisymmetrize the indices $i$ and $j$. The tensor $\xi_k \equiv \frac{1}{2}\e^{ij} N_{i(jk)}$ has one free index and therefore transforms as an $SU(2)$ doublet. Explicitly, its components are
\[\xi_k = \frac{1}{2}(N_{1(2k)}-N_{2(1k)}) = \frac{1}{2}\ml N_{1(21)}-N_{2(11)}\\ N_{1(22)}-N_{2(12)} \mr\;.\]
The tensor $\psi_{ijk} \equiv \frac{1}{2}(N_{i(jk)} + N_{j(ik)})$ is completely symmetric under interchange of the indices $ijk$. Since each index can take on two values, the tensor $\psi_{ijk}$ has only 4 independent components: $\psi_{111},\psi_{112},\psi_{122}$ and $\psi_{222}$. This is the 4 of $SU(2)$, and we have therefore derived the rule $2\x3 = 2\+4$.
\\\\
Let $\Psi \equiv (\psi_{112},\sqrt3\psi_{222},\sqrt3\psi_{111},\psi_{122})$. Under $c$, $\Psi$ transforms as $\Psi \to C\Psi$, where
\[C = \frac{\tau}{2\sqrt2}\ml -1&+\sqrt3&-\sqrt3&+1\\-\sqrt3&-1&+1&+\sqrt3\\ +i\sqrt3&+i&+i&+i\sqrt3\\ +i&-i\sqrt3&-i\sqrt3&+i \mr\;,\]
and under $r$ we have $\Psi \to \text{diag}(+i,+i,-i,-i)\Psi$. Define the change of basis $\Psi \equiv Sq$, where $q \equiv (u,d,c,s)$ and $S$ is a similarity transformation parameterized as
\[S \equiv \ml A&0&B&0\\ \G&0&\Delta&0\\ 0&\al&0&\beta\\ 0&\g&0&\delta \mr\;.\]
Insisting that $S$ makes $C' \equiv S^{-1}CS$ block diagonal fixes the elements of $S$ in terms of $A$ and $B$: $\al = -i\,A,\,\beta = +i\,B,\,\g = -A,\,\del = -B,\,\G = +i\,A,\,\Delta = -i\,B$. The matrix $C'$ is
\begin{equation}\label{eq:generatorsof2primeand2primeprime}
C' = \ml \w\, c&0_{2\times2}\\ 0_{2\times2}&\w^*c \mr
\end{equation}
which implies that $q$ reduces into doublets. Let $\xi' \equiv (u,d)$ and $\xi'' \equiv (c,s)$. Under the operation $c$, evidently we have $\xi' \to c'\xi'$ and $\xi'' \to c''\xi''$, where $c' = \w\,c$ and $c'' = \w^*c$. We have therefore shown that $4 \to 2'\+2''$ when $SU(2)$ is restricted to the subgroup $T'$. Note that $C'^3 = -I$, as it must. Finally, $S^\da S = I$ fixes\footnote{The unitarity of $S$ fixes $|A| = |B| = 1/\sqrt2$ but does not determine the phases of $A$ and $B$. Without loss of generality, we can choose $A$ and $B$ to be real.} $A = B = \frac{1}{\sqrt2}$. Since $q = S^{-1}\Psi$, we have
\[\xi' = \frac{1}{\sqrt2}\ml +\psi_{112}-i\sqrt3\,\psi_{222}\\ -\psi_{122}+i\sqrt3\,\psi_{111} \mr \sim 2'\;\;\text{ and }\;\;\xi'' = \frac{1}{\sqrt2}\ml +\psi_{112}+i\sqrt3\,\psi_{222}\\ -\psi_{122}-i\sqrt3\,\psi_{111} \mr \sim 2''\;.\]
When building Lagrangians based on $T'$, we construct the 4-dimensional representation $\psi_{ijk}$ by multiplying together a $\chi \sim 2$ and a $\phi \sim 2\x_S2 = 3$. In other words, $\psi_{ijk} \equiv -4i\chi_{(i}\phi_{jk)}$, where $\phi_{jk} = \phi_{(jk)}$ and the factor of $-4i$ is just for convenience. We have the doublet $\xi_k \equiv (-4i)\frac{1}{2}\e^{ij}\chi_i\phi_{jk} \sim 2$, and the two doublets $\xi' \sim 2'$ and $\xi'' \sim 2''$ that we just derived. In summary, multiplying a doublet $\chi \sim 2$ and a triplet $\phi \sim 3$ results in $2\x3 = 2\+2'\+2''$, where:
\begin{align}\label{eq:2times3}
&(\chi\phi)_2 = \ml -\sqrt2\phi_+\chi_2-i\,\phi_3\chi_1\\ +\sqrt2\phi_-\chi_1+i\,\phi_3\chi_2 \mr  \nonumber \\
&(\chi\phi)_{2'} = \ml (+\phi_+-i2\sqrt3\phi_-)\chi_2-i\frac{1}{\sqrt2}\phi_3\chi_1\\ (-\phi_-+i2\sqrt3\phi_+)\chi_1+i\frac{1}{\sqrt2}\phi_3\chi_2 \mr \nonumber \\
&(\chi\phi)_{2''} = \ml (+\phi_++i2\sqrt3\phi_-)\chi_2-i\frac{1}{\sqrt2}\phi_3\chi_1\\ (-\phi_--i2\sqrt3\phi_+)\chi_1+i\frac{1}{\sqrt2}\phi_3\chi_2 \mr
\end{align}
with $\phi_\pm \equiv \frac{1}{\sqrt2}(\phi_1\pm i\,\phi_2)$.
\\\\
Earlier we found the transformation properties for a symmetric tensor $M \sim 2\x_S2$ and repackaged its components as a triplet $(M_1,M_2,M_3)$. Similarly, the components of a symmetric tensor $M' \sim 2'\x_S2'$ can be repackaged as the triplet
\begin{equation}\label{eq:2primetimes2prime}
\ml M'_1\\ M'_2\\ M'_3 \mr = \ml -i(M'_{11}+M'_{22})\\ -\w\,(M'_{11}-M'_{22})\\ 2\w^*M'_{12} \mr
\end{equation}
which implies $M'_{11} = \frac{i}{2}(M'_1+i\w\,M_2),\,M_{12} = \frac{1}{2}\w\,M_3$ and $M_{22} = \frac{i}{2}(M_1-i\w\,M_2)$. 
The properties of $M'' \sim 2''\x_S2''$ are found by exchanging $\w \lrarrow \w^*$:
\begin{equation}\label{eq:2primeprimetimes2primeprime}
\ml M''_1\\ M''_2\\ M''_3 \mr = \ml -i(M''_{11}+M''_{22})\\ -\w^*(M''_{11}-M''_{22})\\ 2\w\,M''_{12} \mr
\end{equation}
so $M''_{11} = \frac{i}{2}(M''_1+i\w^*M''_2),\,M''_{12} = \frac{1}{2}\w^*M''_3$ and $M''_{22} = \frac{i}{2}(M''_1-i\w^*M''_2)$.
\\\\
This is all of the group theory required to construct Lagrangians symmetric under the double tetrahedral group. We conclude the appendix with the character table for the double tetrahedral group ($\chi_R \equiv$ character of class in the irreducible representation $R$):
\begin{table}[ht]
\centering
\begin{tabular}{c c c c c c c c c}
\hline\hline
\# of elements in class & class with typical element & $\chi_1$ & $\chi_{1'}$ & $\chi_{1''}$ & $\chi_3$ & $\chi_2$ & $\chi_{2'}$ & $\chi_{2''}$\\ [0.5ex]
\hline
1&$I$&$1$&$1$&$1$&$3$&$2$&$2$&$2$\\
1&$-I$&$1$&$1$&$1$&$3$&$-2$&$-2$&$-2$\\
6&$r,-r$&$1$&$1$&$1$&$-1$&$0$&$0$&$0$\\
4&$c$&$1$&$\w$&$\w^*$&$0$&$1$&$\w$&$\w^*$\\
4&$a$&$1$&$\w^*$&$\w$&$0$&$1$&$\w^*$&$\w$\\
4&$-c$&$1$&$\w$&$\w^*$&$0$&$-1$&$-\w$&$-\w^*$\\
4&$-a$&$1$&$\w^*$&$\w$&$0$&$-1$&$-\w^*$&$-\w$\\
\hline
\end{tabular}
\label{table:chartable}
\end{table}
\\
\textit{Acknowledgments:}
\\\\
This work was completed while one of us (AZ) was visiting the Academia
Sinica in Taipei, Republic of China, whose warm hospitality is greatly
appreciated. This research was partly supported by the NSF under Grant
No. 04-56556.
%\\\\

%\\\\

\begin{thebibliography}{1}
\bibitem{review} M. C. Gonzalez-Garcia and M. Maltoni, ``Phenomenology with Massive Neutrinos," Phys.Rept.460:1-129,2008, arxiv: 0704.1800v2 [hep-ph].
\bibitem{KM} L.-L. Chau and W.-Y. Keung, ``Comments on the Parametrization of the Kobayashi-Maskawa Matrix," Phys. Rev. Lett. 53, 1802 (1984).
\bibitem{jarlskog} C. Jarlskog, ``A Recursive Parameterisation of Unitary Matrices," J.Math.Phys. 46 (2005) 103508m arxiv:math-ph/0504049v3.
\bibitem{2spinor} H. K. Dreiner, H. E. Haber and S. P. Martin, ``Two-component spinor techniques and Feynman rules for quantum field theory and supersymmetry," Physics Reports494:1-196, 2010; Phys.Rept.494:1-196,2010, arXiv:0812.1594v5.
\bibitem{hahn} T. Hahn, ``Routines for the Diagonalization of Complex Matrices," MPP-2006-85, arxiv:physics/0607103v2 [physics.comp-ph].
\bibitem{ma} E. Ma, ``Neutrino Tribimaximal Mixing from $A_4$ Alone," UCRHEP-T472 (August 2009), arxiv:0908.3165v2 [hep-ph].
\bibitem{Tprime} P. H. Frampton and T. W. Kephart, ``Simple Non-Abelian Finite Flavor Groups and Fermion Masses," Int.J.Mod.Phys.A10:4689-4704,1995, arxiv: hep-ph/9409330v1; A. Aranda, C. D. Carone and R. F. Lebed, ``$U(2)$ Flavor Physics without $U(2)$ Symmetry," Phys.Lett. B474 (2000) 170-176, arxiv: hep-ph/9910392v2; M. C. Chen and K. T. Mahanthappa, ``CKM and Tri-bimaximal MNS Matrices in a $SU(5) \times (d)T$ Model," Phys. Lett. B652:34-39, 2007, arxiv:0705.0714 [hep-ph]; F. Feruglio, C. Hagedorn, Y. Lin and L. Merlo, ``Tri-bimaximal Neutrino Mixing and Quark Masses from a Discrete Flavour Symmetry," Nucl.Phys.B775:120-142,2007 (Erratum-ibid.836:127-128,2010) arxiv: hep-ph/0702194v2.
\bibitem{chen} M. C. Chen and K. T. Mahanthappa, ``Geometrical Origin of CP Violation," Phys.Lett.B681:444-447,2009, arxiv: 0904.1721v2 [hep-ph]; M. C. Chen and K. T. Mahanthappa, ``Geometrical Origin of CP Violation and CKM and MNS Matrices in $SU(5) \times T'$," PoS ICHEP2010:407, 2010, arxiv:1011.6364 [hep-ph].
\bibitem{zeeA4} A. Zee, ``Obtaining the Neutrino Mixing Matrix with the Tetrahedral Group," arxiv:hep-ph/0508278v3 4 Oct 2005.
\bibitem{subgroups} W. M. Fairbairn, T. Fulton and W. H. Klink, ``Finite and Disconnected Subgroups of $SU(3)$ and their Application to the Elementary-Particle Spectrum," Journal of Mathematical Physics, Vol. 5 No. 8 pp. 1038-1051, Aug. 1964; K. M. Case, R. Karplus, and C. N. Yang, Phys. Rev. 101, 874 (1956).
\bibitem{wolf} L. Wolfenstein, Phys. Rev. D18, 958 (1978).
\bibitem{hps} P. F. Harrison, D. H. Perkins and W. G. Scott, Phys. Lett. B530, 167 (2002), hep-ph/0202074.
\bibitem{hezee} X. G. He and A. Zee, Phys. Lett. B560, 87 (2003), hep-ph/0301092.
\bibitem{zeeangles} A. Zee, ``Parametrizing the Neutrino Mixing Matrix," Phys.Rev. D68 (2003) 093002, hep-ph/0307323v1 25 Jul 2003.
\end{thebibliography}
\end{document}